\documentclass[prl,reprint,twocolumn]{revtex4-1}
\pdfoutput=1
\usepackage{graphicx}
\usepackage{color}
\usepackage{bm}
\usepackage{hyperref}
\usepackage{amssymb}

\begin{document}

\title{Fractional Quantum Hall Effect and Wigner Crystal of Two-Flux Composite Fermions}
\date{today}

\author{Yang Liu}
\affiliation{Department of Electrical Engineering,
Princeton University, Princeton, New Jersey 08544}
\author{D.\ Kamburov}
\affiliation{Department of Electrical Engineering,
Princeton University, Princeton, New Jersey 08544}
\author{S.\ Hasdemir}
\affiliation{Department of Electrical Engineering,
Princeton University, Princeton, New Jersey 08544}
\author{M.\ Shayegan}
\affiliation{Department of Electrical Engineering,
Princeton University, Princeton, New Jersey 08544}
\author{L.N.\ Pfeiffer}
\affiliation{Department of Electrical Engineering,
Princeton University, Princeton, New Jersey 08544}
\author{K.W.\ West}
\affiliation{Department of Electrical Engineering,
Princeton University, Princeton, New Jersey 08544}
\author{K.W.\ Baldwin}
\affiliation{Department of Electrical Engineering,
Princeton University, Princeton, New Jersey 08544}

\begin{abstract}
  In two-dimensional electron systems confined to GaAs quantum wells,
  as a function of either tilting the sample in magnetic field or
  increasing density, we observe multiple transitions of the
  fractional quantum Hall states (FQHSs) near filling factors
  $\nu=3/4$ and 5/4. The data reveal that these are spin-polarization
  transitions of interacting two-flux composite Fermions, which form
  their own FQHSs at these fillings. The fact that the reentrant
  integer quantum Hall effect near $\nu=4/5$ always develops following
  the transition to full spin polarization of the $\nu=4/5$ FQHS
  strongly links the reentrant phase to a pinned \emph{ferromagnetic}
  Wigner crystal of two-flux composite Fermions.
\end{abstract}


\maketitle

Fractional quantum Hall states (FQHSs) are among the most fundamental
hallmarks of ultra-clean interacting two-dimensional electron systems
(2DESs) at a large perpendicular magnetic field ($B_{\perp}$)
\cite{Tsui.PRL.1982}. These incompressible quantum liquid phases,
signaled by the vanishing of the longitudinal resistance ($R_{xx}$)
and the quantization of the Hall resistance ($R_{xy}$), can be
explained by mapping the interacting electrons to a system of
essentially non-interacting, 2$p$-flux composite Fermions
($^{2p}$CFs), each formed by attaching $2p$ magnetic flux quanta to an
electron ($p$ is an integer).  The $^{2p}$CFs have discrete energy
levels, the so-called $\Lambda$-levels, and the FQHSs of electrons
seen around Landau level (LL) filling factor $\nu=1/2$ (1/4) would
correspond to the integer quantum Hall states of $^2$CFs ($^4$CFs) at
integral $\nu^{CF}$ \cite{Jain.CF.2007}. In state-of-the-art,
high-mobility 2DESs, FQHSs also develop around $\nu=3/4$, and are
usually understood as the particle-hole counterparts of the FQHSs near
$\nu=1/4$ through the relation $\nu\leftrightarrow (1-\nu$) \footnote{In
  spinless or fully spin-polarized systems, the interacting electrons
  at $\nu$ are equivalent to interacting holes at $(1-\nu)$. In spinful
  systems, each LL is 2-fold degenerate and the particle-hole symmetry
  relates $\nu$ to $(2-\nu)$.}
\cite{Yeh.PRL.1999}. Alternatively, these states might also be the
FQHSs of \emph{interacting} $^{2}$CFs at $\nu^{CF}=\nu/(1-2\nu)$. For
example, the $\nu=4/5$ state is the $\nu^{CF}=-4/3$ FQHS of
$^{2p}$CFs, and has the same origin as the unconventional FQHS seen at
$\nu=4/11$ ($\nu^{CF}=4/3$) \cite{Pan.PRL.2003, Chang.PRL.2004}.

\begin{figure}[htbp]
\includegraphics[width=.45\textwidth]{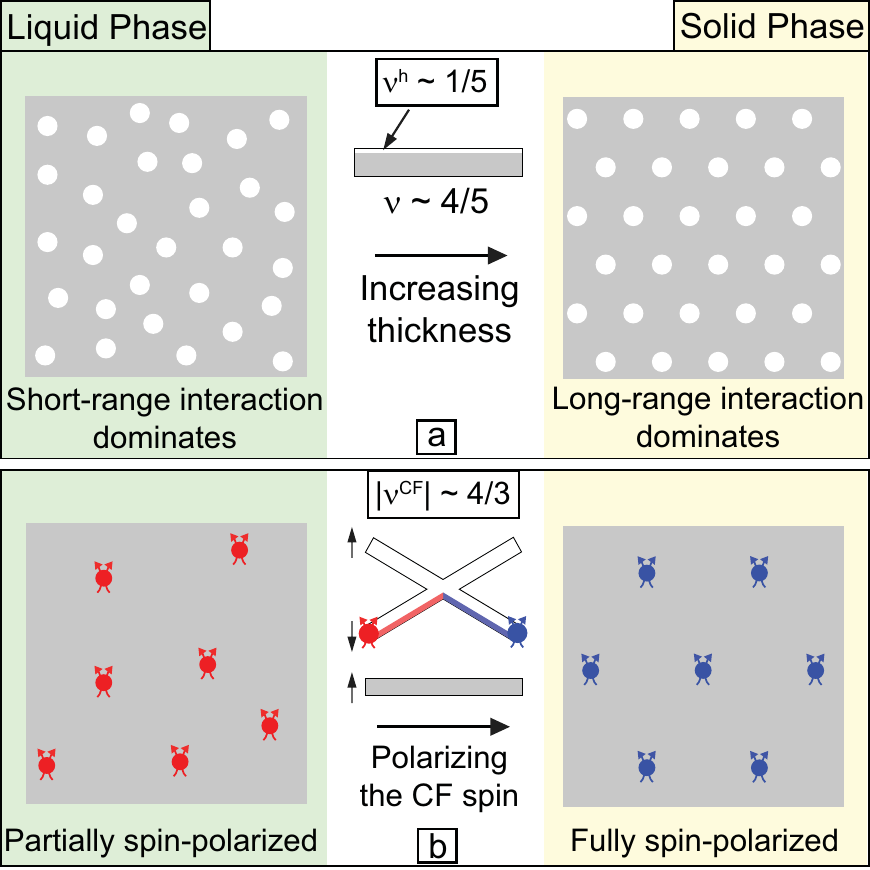}
\caption{Schematic figures illustrating two different explanations of
  the RIQHS near $\nu=4/5$. (a) The hole Wigner crystal picture. The
  electrons at $\nu\sim 4/5$ are equivalent to holes at $\nu^h\sim
  1/5$. These holes condense into a liquid phase when the short-range
  interaction dominates (left), and into a crystal phase when the
  long-range interaction dominates in thicker 2DESs (right). The gray
  background represents electrons, and white circles represent the
  holes. (b) The $^2$CF Wigner crystal picture. The electrons at
  $\nu\sim 4/5$ are equivalent to $^2$CFs at $|\nu^{CF}|\sim
  4/3$. There is one fully-filled, spin-up $\Lambda$-level (the gray
  background), and the rest of the $^2$CFs can either be spin-down
  (red) and form a liquid phase when the Zeeman energy ($E_Z$) is
  small (left panel), or be spin-up (blue) at large $E_Z$ and form a
  ferromagnetic crystal phase of $^2$CFs (right panel).}
\end{figure}

Another hallmark of clean 2DESs is an insulating phase that terminates
the series of FQHSs at low fillings, near $\nu=1/5$,
\cite{Jiang.PRL.1990, Goldman.PRL.1990}. This insulating phase is
generally believed to be an electron Wigner crystal, pinned by the
small but ubiquitous disorder potential \footnote{See articles by
  H. A. Fertig and by M. Shayegan, in \emph{Perspectives in Quantum
    Hall Effects}, Edited by S. Das Sarma and A. Pinczuk (Wiley, New
  York, 1997).}. Recently, an insulating phase was observed near
$\nu=4/5$ in clean 2DESs \cite{Liu.PRL.2012,
  Hatke.Nat.Comm.2014}. This phase, which is signaled by a reentrant
integer quantum Hall state (RIQHS) near $\nu=1$, was interpreted as
the particle-hole symmetric state of the Wigner crystal seen at very
small $\nu$ \cite{Liu.PRL.2012, Hatke.Nat.Comm.2014}. In this picture,
the holes, unoccupied states in the lowest LL, have filling factor
$\nu^h\sim 1/5$ ($=1-4/5$) and form a liquid phase when the
short-range interaction is strong; see the left panel of
Fig. 1(a). They turn into a solid phase when the thickness of the 2DES
increases and the long-range interaction dominates (right panel of
Fig. 1(a)). This interpretation is plausible, since the RIQHS only
appears when the well-width ($W$) is more than five times larger than
the magnetic length. However, it does not predict or allow for any
transitions of the $\nu=4/5$ FQHS, which is always seen just before
the RIQHS develops \cite{Liu.PRL.2012}.

Here we report our extensive study of the FQHSs near $\nu=3/4$ (at 4/5
and 5/7) and their particle-hole counterparts near $\nu=5/4$ (at 6/5
and 9/7) \cite{Note1}. Via either increasing the 2DES density or the
tilt angle between the magnetic field and the sample normal, we
increase the ratio of the Zeeman energy ($E_Z=|g|\mu_BB$ where $B$ is
the total magnetic field) to Coulomb energy ($V_C=e^2/4\pi\epsilon
l_B$ where $l_B=\sqrt{\hbar/e B_{\perp}}$ is the magnetic length), and
demonstrate that these FQHSs undergo multiple transitions as they
become spin polarized. The number of observed transitions, one for the
FQHSs at $\nu=4/5$ and 6/5, and two for the states at $\nu=5/7$ and
9/7, is consistent with what is expected for polarizing the spins of
\textit{interacting} $^2$CFs. Note that these interacting $^2$CFs form
FQHSs at \textit{fractional} CF fillings $\nu^{CF}=-4/3$ ($\nu=4/5$
and 6/5) and -5/3 ($\nu=5/7$ and 9/7). Even more revealing is the
observation that, whenever the RIQHS near $\nu=4/5$ develops, it is
preceded by a transition of the FQHS at $\nu=4/5$ to a fully
spin-polarized $^2$CF state. This provides evidence that the RIQHS is
the manifestation of a \textit{ferromagnetic $^2$CF Wigner crystal}
(see Fig. 1(b)).


We studied 2DESs confined to wide GaAs quantum
wells (QWs) bounded on each side by undoped Al$_{0.24}$Ga$_{0.76}$As
spacer layers and Si $\delta$-doped layers, grown by molecular beam
epitaxy. We report here data for two samples, with QW widths $W=$ 65
and 60 nm, and as-grown densities of $n\simeq$ 1.4 and 0.4, in units
of $10^{11}$ cm$^{-2}$ which we use throughout this report. The
samples have a van der Pauw geometry with InSn contacts at their
corners, and each is fitted with an evaporated Ti/Au front-gate and an
In back-gate. We carefully control $n$ while keeping the charge
distribution symmetric. The measurements were carried out in dilution
refrigerators with a base temperature of $T \simeq$ 25 mK and
superconducting magnets. We used low-frequency ($\lesssim$ 40 Hz)
lock-in techniques to measure the transport coefficients.

\begin{figure}
\includegraphics[width=.45\textwidth]{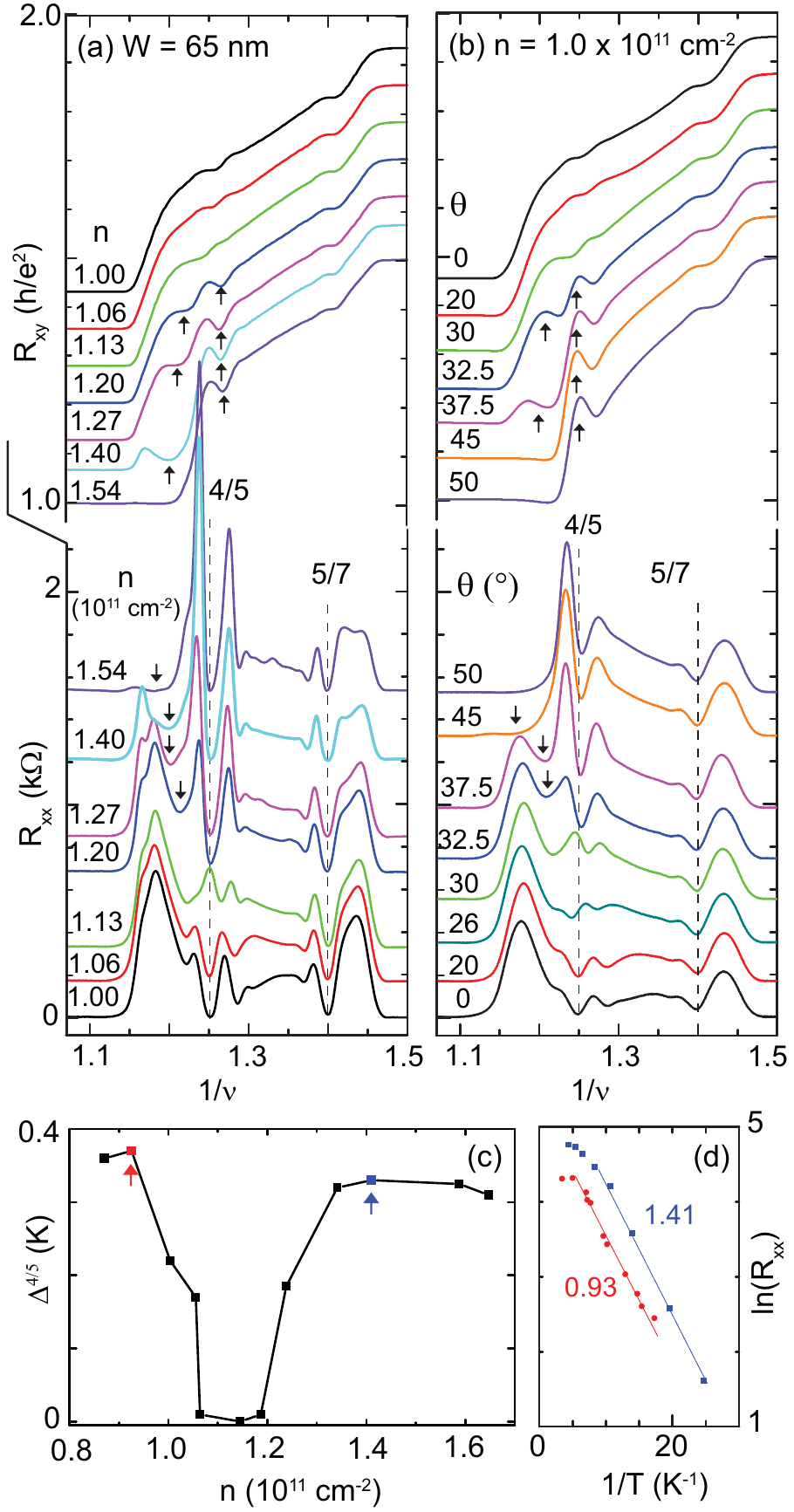}
\caption{Longitudinal ($R_{xx}$) and Hall ($R_{xy}$)
  magnetoresistance traces for 2D electrons confined to a 65-nm-wide
  GaAs QW near $\nu=3/4$ as a function of (a) increasing charge
  density, and (b) tilting the sample in the magnetic field. The
  density $n$ (in units of $10^{11}$ cm${^{-2}}$) or tilting angle
  $\theta$ for each trace is indicated, and traces are shifted
  vertically for clarity. (c) Energy gap of the
  $\nu=4/5$ FQHS as a function of $n$. (d) Arrehnius plot of
  $R_{xx}$ vs. $1/T$ at $n=0.93$ and $1.41\times 10^{11}$ cm$^{-2}$.}
\end{figure}

\begin{figure*}[htbp]
\includegraphics[width=.8\textwidth]{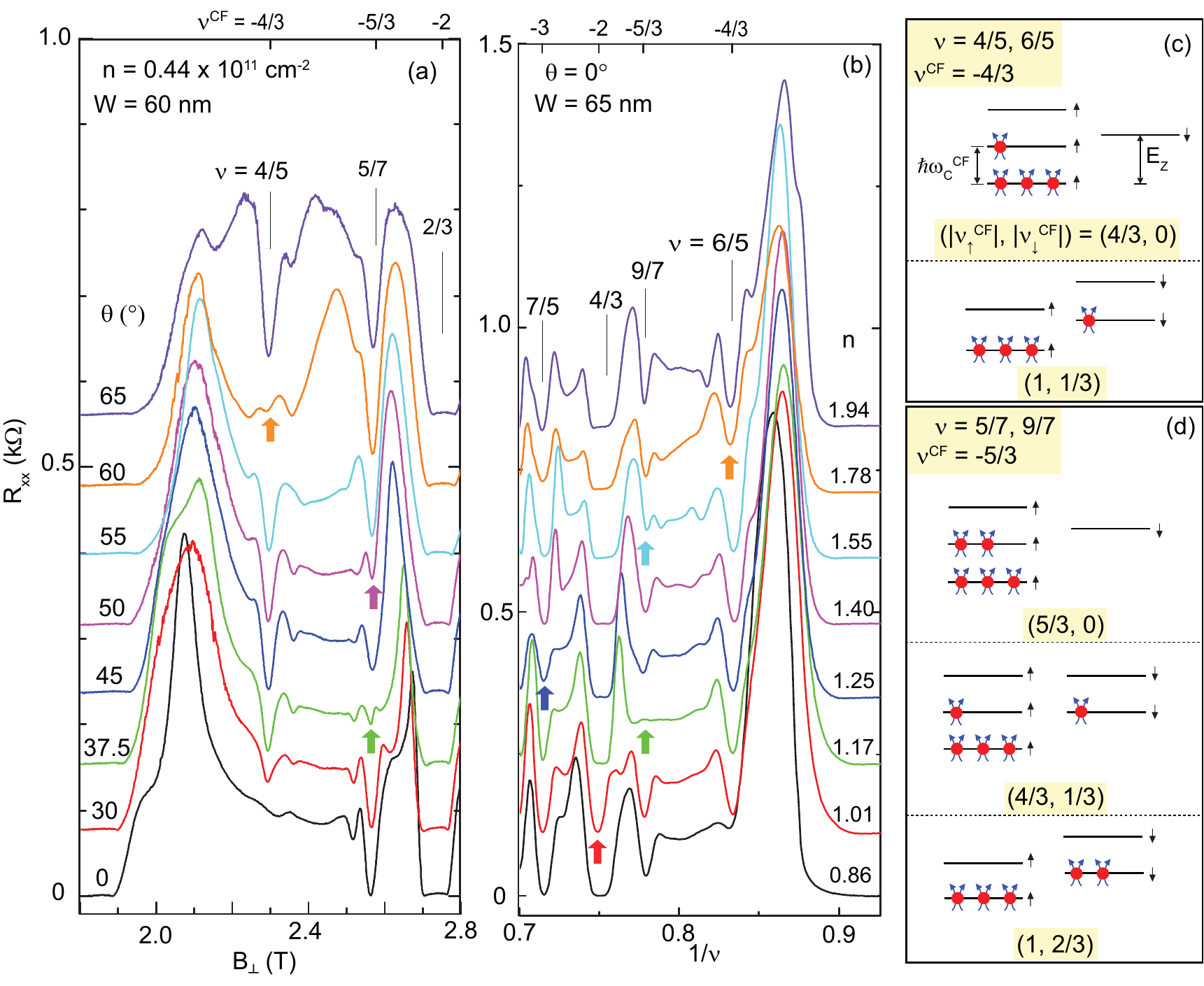}
\caption{(a) $R_{xx}$ measured in a 60-nm-wide QW near $\nu=3/4$, at a
  fixed density $n=0.44\times 10^{11}$ cm$^{-2}$ and different tilting
  angles $\theta$. (b) $R_{xx}$ for a 65-nm-wide QW near $\nu=5/4$, at
  $\theta=0^{\circ}$ and different densities $n=0.86$ to $1.94\times
  10^{11}$ cm$^{-2}$. (c, d) Schematic plots showing multiple
  configurations of the $\nu=4/5$ and 6/5, and 5/7 and 9/7 FQHSs with
  different spin-polarizations.}
\end{figure*}

\begin{figure}[htbp]
\includegraphics[width=.45\textwidth]{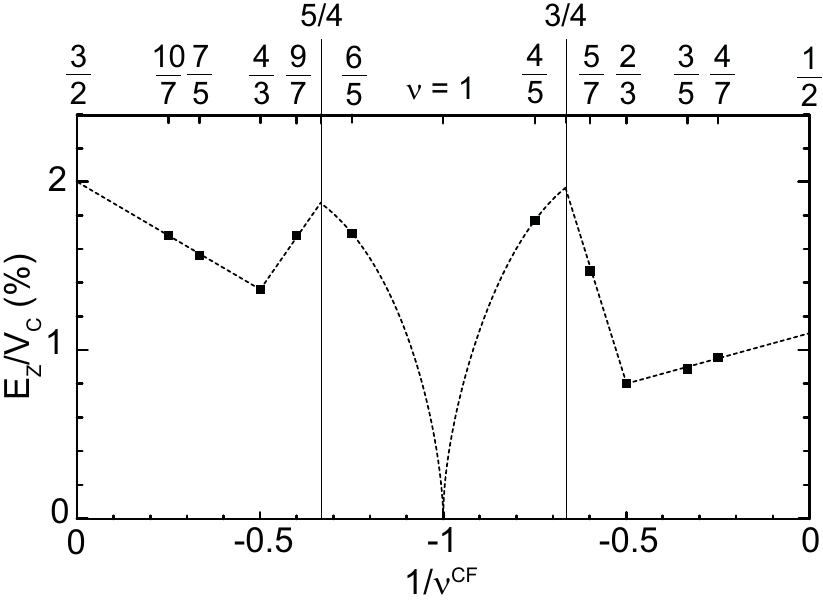}%
\caption{Summary of the spin-polarization energy in units of the
  Coulomb energy, $E_Z/V_C$, at different filling factors
  ($\nu$). Dotted lines are guides to the eye. All data points were
  measured in symmetric QWs. The transitions at $\nu=2/3$, 3/5 and 4/7
  were measured in perpendicular magnetic field by changing $n$, and
  the rest at a fixed density $n=0.44$ by changing $\theta$. For each
  filling, only the (last) transition into a fully spin-polarized
  configuration is shown.}
\end{figure}

Figure 2(a) shows $R_{xx}$ and $R_{xy}$ magnetoresistance traces near
$\nu=3/4$ measured in a symmetric 65-nm-wide QW, at densities ranging
from 1.00 to 1.54. The deep $R_{xx}$ minimum seen at $\nu=4/5$ in the
lowest density ($n=1.00$) trace disappears at $n=1.13$ and reappears
at higher densities. With increasing $n$, an $R_{xx}$ minimum also
develops to the left of $\nu=4/5$, and merges with the $\nu=1$
$R_{xx}=0$ plateau at the highest density $n=1.54$ (see down arrows in
Fig. 2(a)). Meanwhile, two minima appear in $R_{xy}$ on the sides of
$\nu=4/5$ when the $\nu=4/5$ FQHS reappears (up-arrows in
Fig. 2(a)). These two $R_{xy}$ minima become deeper at higher
densities and, at $n\simeq 1.54$, the $R_{xy}$ minimum on the left
side of $\nu=4/5$ merges into the $\nu=1$ $R_{xy}=h/e^2$ plateau
\footnote{As demonstrated in Ref. \cite{Liu.PRL.2012} (see, e.g.,
  Fig. 3(c) of Ref. [10]), at the lowest temperatures, $R_{xy}$
  minimum to the right of $\nu=4/5$ also approaches the $\nu=1$
  plateau ($=h/e^2$), and $R_{xy}$ at 4/5 becomes quantized at
  $(5/4)(h/e^2)$.}.

The $R_{xx}$ and $R_{xy}$ data of Fig. 2(a) provide evidence for the
development of a RIQHS between $\nu=4/5$ and 1, as reported recently
and attributed to the formation of a pinned Wigner crystal state
\cite{Liu.PRL.2012, Hatke.Nat.Comm.2014}. Note that at the onset of
this development, the $\nu=4/5$ FQHS shows a transition manifested by
a weakening and strengthening of its $R_{xx}$ minimum. The central
questions we address here are: What is the source of this transition,
and what does that imply for the origin of the $\nu=4/5$ FQHS and the nearby RIQHS?

As Fig. 2(b) illustrates, the $R_{xx}$ and $R_{xy}$ traces measured in
the same QW at a fixed density $n = 1.00$ and different tilting angles
$\theta$ reveal an evolution very similar to the one seen in Fig. 2(a)
($\theta$ denotes the angle between the magnetic field and normal to
the 2D plane). At $\theta=0^{\circ}$, a strong FQHS is seen at
$\nu=4/5$. The $R_{xx}$ minimum at $\nu=4/5$ disappears at
$\theta\simeq 30^{\circ}$ and reappears at higher $\theta$, signaling
the destruction and resurrection of the FQHS. Two minima in $R_{xy}$
on the sides of $\nu=4/5$, marked by the up-arrows, develop at
$\theta>30^{\circ}$. As $\theta$ is further increased, the $R_{xy}$
minimum to the left of $\nu=4/5$ deepens and an $R_{xx}$ minimum
starts to appear at the same $\nu$ (see down-arrows in Fig. 2(b)). At
the highest tilting angles $\theta>40^{\circ}$, these minima merge
into the $\nu=1$ $R_{xy}$ Hall plateau quantized at $h/e^2$, and the
$R_{xx}=0$ minimum near $\nu=1$ plateau, respectively. This evolution
with increasing $\theta$ is clearly very similar to what is seen as a
function of increasing electron density. Moreover, it suggests that
the $\nu=4/5$ FQHS transition is induced by the enhancement of $E_Z$
and is therefore spin related, similar to the spin-polarization
transitions observed for other FQHSs \cite{Clark.PRL.1989,
  Eisenstein.PRL.1989, Du.PRL.1995, Liu.cond.mat.2014}.

Observing a spin-polarization transition for the $\nu=4/5$ FQHS,
however, is surprising as this state is usually interpreted as the
particle-hole counterpart of the $\nu=1/5$ FQHS, which is formed by
\emph{non-interacting} four-flux $^4$CFs. Such a state should be
always fully spin-polarized and no spin-polarization transition is
expected \cite{Yeh.PRL.1999}. On the other hand, if the $\nu=4/5$ FQHS
is interpreted as the FQHS of \emph{interacting} two-flux CFs
($^2$CFs), then it corresponds to the $\nu^{CF}=-4/3$ FQHS of $^2$CFs,
and has two possible spin configurations as shown in Fig. 3(c). The
system has one fully-occupied, spin-up, $\Lambda$-level and one
1/3-occupied $\Lambda$-level. Depending on whether $E_Z$ is smaller or
larger than the $\Lambda$-level separation of the $^2$CFs, the
1/3-filled $\Lambda$-level may be either spin-down or spin-up (see
Fig. 3(c)) \footnote{When $E_Z/V_C$ is extremely small, the $\nu=4/5$
  FQHS can possibly have a third, spin-singlet configuration where the
  spin-up and -down $\Lambda$-levels are both 2/3-occupied.}.

  To further test the validity of the above interpretation, we
  measured $R_{xx}$ in a 60-nm-wide QW at a very low density $n=0.44$
  and different $\theta$ in Fig. 3(a). The $\nu=4/5$ FQHS exhibits a
  clear transition at $\theta=60^{\circ}$, manifested by a weakening
  of the $R_{xx}$ minimum. Note that the transition of the $\nu=4/5$
  FQHS appears in Figs. 2(a), 2(b) and 3(a) when the ratio of the
  Zeeman to Coulomb energies ($E_Z/V_C$) is about 0.0145, 0.0157 and
  0.0177, respectively. The electron layer-thicknesses at these three
  transitions, parameterized by the standard deviation ($\lambda$) of
  the charge distribution in units of $l_B$, are 1.66, 1.52 and 0.75,
  respectively. The softening of the Coulomb interaction due to the
  finite-layer-thickness effect is less and the spin-polarization
  energy should be higher for smaller $\lambda/l_B$ (see
  \cite{Liu.cond.mat.2014} for the dependence of spin-polarization
  energy on the finite-layer-thickness). Therefore, these values are
  consistent with each other.

In Fig. 3(a), we also observe two transitions for the $\nu=5/7$ FQHS
at $\theta=37.5^{\circ}$ and $50^{\circ}$, suggesting three different
phases. This observation is consistent with the $\nu=5/7$ FQHS being
formed by interacting $^2$CFs. In such a picture, the $\nu=5/7$ FQHS,
which is the $\nu^{CF}=-5/3$ FQHS of the $^2$CFs, has three different
possible spin configurations, as shown in Fig. 3(d). Similar to
Fig. 3(c), the lowest spin-up $\Lambda$-level is always fully
occupied. The second $\Lambda$-level is 2/3-occupied spin-up
(spin-down), if $E_Z$ is larger (smaller) than the $\Lambda$-level
separation. If $E_Z$ equals the $\Lambda$-level separation, the
$^2$CFs form a novel spin-singlet state when the spin-up and spin-down
$\Lambda$-levels are both 1/3-occupied; see the middle panel of
Fig. 3(d).

Data near $\nu=5/4$ measured in the 65-nm-wide QW at different
densities, shown in Fig. 3(b), further confirm our picture. The
$\nu=6/5$ and 9/7 FQHSs exhibit transitions similar to their
particle-hole conjugate states at $\nu=4/5$ and 5/7, respectively. The
$\nu=6/5$ FQHS shows a transition at $n=1.78$. At this transition,
$E_Z/V_C\simeq 0.0149$ and $\lambda/l_B\simeq 1.86$, very similar to
the corresponding values (0.0145 and 1.66) at $\nu=4/5$ in Fig. 2(a),
suggesting that the particle-hole symmetry $\nu\leftrightarrow
(2-\nu)$ is conserved in this case \footnote{In
  Ref. \cite{Liu.cond.mat.2014}, the FQHSs with integer $\nu^{CF}$
  near $\nu=1/2$ and 3/2 have very different spin polarization
  energies at similar $\lambda/l_B$, suggesting that the particle-hole
  symmetry is broken.}. Furthermore, the $\nu=9/7$ FQHS becomes weak
twice, at $n=1.17$ and 1.55, also consistent with the $\nu=5/7$ FQHS
transitions.



It is instructive to compare the transitions we observe at fractional
$\nu^{CF}$ with the spin-polarization transitions of other FQHSs at
integer $\nu^{CF}$. In Fig. 4, we summarize the critical $E_Z/V_C$
above which the FQHSs between $\nu=1/2$ and 3/2 become fully
spin-polarized. The measurements were all made on the 60-nm-wide
QW. The $x$-axis is $1/\nu^{CF}$, and we mark the electron LL filling
factor $\nu$ in the top axis. The dotted lines, drawn as guides to the
eye, represent the phase boundary between fully
spin-polarized (above) and partially spin-polarized (below)
$^2$CFs. Note that the system is always fully spin-polarized at
$\nu^{CF}=-1$ ($\nu=1$) \cite{Kukushkin.PRL.1999,
  Tiemann.Science.2012}. The critical $E_Z/V_C$ of FQHSs with integral
$\nu^{CF}$ increases with $\nu^{CF}$ and reaches maxima at
$\nu^{CF}=-\infty$ ($\nu=1/2$ and 3/2). Secondary maxima in the
boundaries appear at $\nu^{CF}=-3/2$ ($\nu=3/4$ and 5/4), and seem to
have approximately the same height as at $\nu=1/2$ and 3/2.

While Fig. 3 data strongly suggest that we are observing spin
transitions of various FQHSs, there is also some theoretical
justification. It has been proposed that the enigmatic FQHSs observed
at $\nu=4/11$ and 5/13 in the highest quality samples can be
interpreted as the FQHSs of interacting $^2$CFs at $\nu^{CF}=+4/3$ and
+5/3 \cite{Pan.PRL.2003, Chang.PRL.2004}. A recent theoretical study
predicts a transition of the $\nu=4/11$ FQHS to full spin polarization
when $E_Z/V_C$ is about 0.025 \cite{Mukherjee.PRL.2014}. Our observed
transition of the $\nu=4/5$ ($\nu^{CF}=-4/3$) FQHS appears at $E_Z/V_C
\simeq 0.015$ to 0.024, in different QWs with well width ranging from
65 to 31 nm and corresponding $\lambda/l_B \simeq 1.7$ to 1.1
\cite{Liu.PRL.2012}, consistent with this theoretically predicted
value.

Another useful parameter in characterizing the origin of the $\nu=4/5$
FQHS and its transition are the energy gaps on the two sides of the
transition. We show in Fig. 2(c) the measured excitation gaps for this
state at different densities in the 65-nm-wide QW. The $\nu=4/5$ FQHS
transition for this sample occurs at density $n \simeq 1.1$, see
Fig. 2(a). Before and after the transition, e.g. at $n=0.86$ and 1.64,
the $\nu=4/5$ FQHS has very similar energy gaps ($\sim 0.35$ K)
although the densities are different by nearly a factor of two. Since
the FQHS energy gaps at a given filling are ordinarily expected to
scale with $V_C \sim \sqrt{n}$), this observation suggests that the
excitation gap at $\nu=4/5$ is reduced when the FQHS becomes fully
spin polarized.

Finally we revisit the RIQHSs we observe near $\nu=4/5$ (see, e.g.,
Figs. 1(a) and 1(b)). These RIQHSs were
interpreted as pinned Wigner crystal states \cite{Liu.PRL.2012}, and
the recent microwave resonance experiments confirm this interpretation
\cite{Hatke.Nat.Comm.2014}. Moreover, the data in
Ref. \cite{Liu.PRL.2012} as well as the data we have presented here
all indicate that, whenever a transition to a RIQHS occurs, it is
initiated by a transition of the $\nu=4/5$ FQHS. As we have
shown here, the transition we see for the $\nu=4/5$ FQHS is a
transition to a \textit{fully spin polarized state of interacting
  $^2$CFs}. Combining these observations leads to a tantalizing
conclusion: The fact that the RIQHS near $\nu=4/5$ always develops
following a spin polarization transition of $^2$CFs strongly links the
RIQHS to a pinned, \textit{ferromagnetic Wigner crystal of $^2$CFs}, as
  schematically illustrated in Fig. 1(b).

\begin{acknowledgments}

  We acknowledge support through the NSF (DMR-1305691) for
  measurements, and the Gordon and Betty Moore Foundation (Grant
  GBMF2719), Keck Foundation, the NSF MRSEC (DMR-0819860), and the DOE
  BES (DE-FG02-00-ER45841) for sample fabrication. A portion of this
  work was performed at the National High Magnetic Field Laboratory,
  which is supported by the NSF (Cooperative Agreement
  No. DMR-1157490), State of Florida, and DOE. We thank S. Hannahs,
  G. E. Jones, T. P. Murphy, E. Palm, and J. H. Park for technical
  assistance, and J. K. Jain for illuminating discussions.

\end{acknowledgments}

\bibliography{../bib_full}
\end{document}